# Control over emissivity of zero-static-power thermal emitters based on phase changing material GST

Kaikai Du[1], Qiang Li[1], Yanbiao Lyu[2], Jichao Ding[1], Yue Lu[1], Zhiyuan Cheng[2] and Min Qiu[1]

[1]State Key Laboratory of Modern Optical Instrumentation, College of Optical Science and Engineering, Zhejiang University, 310027, Hangzhou, China

[2]College of Information Science & Electronic Engineering, Zhejiang University, 310027, Hangzhou, 310027, China

E-mail: qiangli@zju.edu.cn

**Abstract:**

Controlling the emissivity of a thermal emitter has attracted growing interest with a view towards a new generation of thermal emission devices. So far, all demonstrations have involved sustained external electric or thermal consumption to maintain a desired emissivity. Here control over the emissivity of a thermal emitter consisting of a phase changing material $Ge_2Sb_2Te_5$ (GST) film on top of a metal film is demonstrated. This thermal emitter shows broad wavelength-selective spectral emissivity in the mid-infrared. The peak emissivity approaches the ideal blackbody maximum and a maximum extinction ratio of above 10dB is attainable by switching GST between the crystalline and amorphous phases. By controlling the intermediate phases, the emissivity can be continuously tuned. This switchable, tunable, wavelength-selective and thermally stable thermal emitter will pave the way towards the ultimate control of thermal emissivity in the field of fundamental science as well as for energy-harvesting and thermal control applications, including thermophotovoltaics, light sources, infrared imaging and radiative coolers.

**Introduction**

Any object with a temperature above absolute zero gives off light or thermal emission. Black soot generally shows a high emissivity while smooth metal exhibits a low emissivity. Other natural materials have certain emissivities as well. The intrinsic thermal emission of these materials is usually spectrally broad and their emissivities cannot be easily tuned. However, in some applications such as biochemical sensing,[1] light sources[2, 3] thermal emitters with tailored emissivities are highly desirable. For instance, a sharp cut-off in emission wavelength is needed in thermophotovoltaics;[4-8] passive radiative coolers demand the emitter to radiate only in the spectral range of 8-13 μm.[9-11] Besides, switchable and tunable thermal emission is necessary in applications such as infrared camouflage[12] and heat management.[13-15]

Conventional thermal emission with spectral selectivity is based on micro/nano-structures such as gratings,[16] photonic crystals,[17] photonic cavities,[18-20] nanoantennas[21] and metamaterials.[22] By altering structural parameters of the micro/nano-structures with advanced micro/nano-fabrication, wavelength-selective thermal emitters from THz to visible can be realized.[16-30] However, the static micro/nano-structures does not offer flexible tunability in thermal emissivity. So far several strategies have been implemented to engineer control over the thermal emissivity. (1) Electrically tunable materials (such as quantum wells and graphene) have been employed for modulation of the thermal emissivity. By electrically tuning intersubband absorption in n-type quantum wells, the emissivity of a narrow-band mid-infrared thermal emitter based on a photonic crystal slab can be tailored statically[31, 32] or dynamically.[33-35] Using an electrostatic gate to control the charge density in graphene, modulation of the emissivity of a narrow-band mid-infrared thermal emitter composed of graphene/SiNx/Au nanoresonators by 3% has been achieved.[36] However, sustained electric consumption is in need to maintain a certain state of the tunable materials. (2) Materials with high thermal expansion coefficients (such as SiNx) have been introduced in micro-electro-mechanical-system based emitters.[37] By manipulating the distance between the top metamaterial pattern and the bottom metallic film through applying heat, control of the temperature-dependent emissivity is attained. However, the emissivity difference is lower than 35% and sustained thermal consumption needs to be applied to keep the mechanical distance so as to control the emissivity. (3) Phase changing materials (PCMs) have also been introduced to modulate the thermal emissivity owing to different optical and infrared properties in different phases. VOx is a typical PCM and exhibits an insulator-to-metal phase transition at a temperature of 67 ℃.[12] By controlling the phases of VOx at different temperatures, a tunable thermal emissivity for an emitter composed of a VOx layer on a sapphire substrate has been demonstrated.[12] However, to preserve the metal phase of VOx, the temperature of the thermal emitter needs to be maintained above 67 ℃. Otherwise, a metal-to-insulator phase recovery is activated. Although significant progress has been made to control the thermal emissivity, there has yet to emerge an energy-efficient solution to maintain a desired emissivity with zero static power consumption at room temperature.

$Ge_2Sb_2Te_5$ (GST) is another PCM material with amorphous and crystalline phases.[38-45] It has been widely used as a commercialized rewriteable optical storage medium in digital video disks owing to its phase changing performance.[46-47] The atom distributions of GST in amorphous and crystalline phases are shown in Fig. 1a.[48] While the atoms are chaotically arranged in the amorphous phase, they are aligned orderly in the crystalline phase, leading to distinct infrared properties. The GST film can be fabricated by magnetron sputtering and the as-deposited GST film is initially at the amorphous phase. The amorphous GST (termed as "aGST") transforms to the crystalline phase (termed as "cGST") when it is annealed above 160°C.[49] For reamorphization, it needs a quick annealing process over 640°C.[50] Once the crystallization or reamorphization has been accomplished, the GST film stays stable in a certain phase even for years at room temperature unless the phase transition conditions are satisfied, making it a very appealing material for energy-efficient optoelectronics devices with zero static power. Based on these unique properties, GST has been applied in various energy-efficient switchable photonic devices such as absorbers,[39] multi-level memory,[51] chiral metamaterials,[40] and color devices.[41, 52, 53]

In this paper, control over the emissivity of a zero-static-power mid-infrared thermal emitter based on a GST film on top of a metal film is demonstrated. The emissivity of this thermal emitter is switchable, tunable and wavelength-selective. The wavelength-selectivity is accomplished by altering the GST thickness. The switchable thermal emission is achieved by switching GST between amorphous and crystalline phases. By controlling the intermediate phases composed of different proportions of amorphous and crystalline molecules of GST, the emissivity of the thermal emitter can be continuously tuned. This GST based thermal emitter presents several distinctive advantages. (1) The two phases (amorphous and crystalline) of GST show distinct infrared properties, resulting in a sharp contrast between on and off states in the switchable emissivity. When the GST is at the crystalline phase, an emissivity approaching the ideal blackbody maximum (on state) is realized while the emissivity is significantly suppressed to below 0.2 (off state) at the amorphous phase. By switching between amorphous and crystalline phases of GST, a maximum extinction ratio up to

11 dB can be realized. (2) The transition between amorphous and crystalline phases of GST reveals hybrid phase behaviors, which can be exploited to implement tunable emission by properly controlling annealing time and annealing temperature. (3) Wavelength-selective thermal emission in a broad range (from 3 μm to 15 μm) can also be enabled through changing the GST thickness. (4) The fabrication of this thermal emitter involves only simple film deposition; therefore, this thermal emitter presents the virtue of large-area and lithography-free fabrication as well as design flexibility.

**Materials and Methods**

*Fabrication of the GST-Au samples:* The GST-Au samples are fabricated starting with a 120-nm-thick gold film deposition. As the gold film is optically thick enough, there is low demand for the substrate. The gold film is deposited on the BK7 glass substrate by thermal evaporation technique. The GST layer is then deposited on the gold film by magnetron sputtering, in which germanium is DC sputtered and the other two components (stibium and tellurium) are RF sputtered with a 2:2:5 deposition rate ratio. The as-deposited GST alloy is at the amorphous phase and an annealing process at 180 ℃ on a hot plate for 1.5 min is applied to get the crystalline GST in this paper.

*Absorptivity measurements:* The asborptivities of GST-Au samples are derived by measuring the reflectivities. As the gold film is optically thick enough, transmission doesn't exist in the experiment. Thus the absorptivity ($A$) and the reflectivity ($R$) satisfy $A=1-R$. The vertical reflectivities are measured with the Bruker Vertex 70 FTIR and Hyperion 1000 infared microscope equipped with a liquid nitrogen refrigerated MCT detector and the oblique reflectivities are measured with a DTGS detector on Vertex 70 FTIR equipped with the A513 attachment.

*Emissivtity measurements:* The emitted spectra of the black soot and the GST-Au samples are measured by the Bruker Vertex 70 FTIR equipped with a DTGS detector.

The samples are fixed on a temperature controller and the emitted power is sent into the FTIR and detected by the DTGS detector. The temperatures of the black soot and the GST-Au samples are controlled over 100 ℃ to ensure high emitted power and decrease the noise. Every measurement is repeated 16 times to further reduce the noise. The infared photoographs are taken by the FLIR TG165 infrared camera.

**Results and Discussion**

A schematic diagram of the switchable and tunable thermal emitter is shown in Fig. 1b. A GST layer is deposited on a gold film (also termed as "GST-Au sample"). The gold film, which is optically thick enough, can be replaced by other metal (silver, aluminum, etc) films. Figure 1c presents an SEM image of a cross-section of the device with a 450-nm-thick GST on top of a 120-nm-thick gold layer.

According to Kirchhoff's law of thermal radiation, the absorptivity of an object is equal to its thermal emissivity. Before revealing the emissivity of the thermal emitter, its absorption properties as an absorber are studied first. The electromagnetic responses of the GST-Au samples are simulated by COMSOL Multiphysics software. In the simulation, the relative permittivity of gold is derived from Rakić's work.[54] The relative permittivity of GST (shown in Fig. S1) is determined based on spectrophotometric method from the transmission and reflection measurements.[55] From the measured permittivity, it can be concluded that GST at the amorphous phase is transparent in the mid-infrared while GST at the crystalline phase is highly absorptive by contrast.

The absorptivities of GST-Au samples at normal incidence are investigated by simulations and experiments, which are presented in Fig. 2a and Fig. 2b, respectively. The absorptivities of the samples at oblique incidence are presented in Fig. S2. In the mid-infrared, the metal layer can be approximately regarded as a perfect electric conductor (PEC) due to its large real and imaginary parts of refractive index. Thus, nearly complete reflection with a phase shift of $\pi$ at the GST-Au interface is formed. Without the metal layer, fundamental resonance can also occurs at a GST thickness of around $\lambda/2n$ ($n$ is the real part of the refractive index of GST and $\lambda$ represents the resonant wavelength); however, the maximum absorption is limited to 50%[56, 57] (Fig. S3). When a gold film is placed below the GST film, not only the GST thickness is halved, but also a nearly unity absorptivity can be achieved for the cGST-Au sample (see Section 3 of the Supporting Information for more detail). As aGST is nearly transparent, a resonant cavity is formed at a GST thickness of around $\lambda/4n$ so that a round trip inside the cavity could accumulate a 0 (or $2\pi$) phase shift.[58] At resonance, there is still absorption and field penetration in the metal film in both aGST-Au and cGST-Au samples as the metal film isn't an ideal PEC. For the aGST-Au sample, this little absorption and field penetration in the metal film contribute to the total absorption (below 20%) at resonance; while for the cGST-Au sample, nearly unity absorption arises from resonance induced field enhancement in the highly absorptive cGST layer. Therefore, a sharp contrast between absorptivities of aGST-Au and cGST-Au samples is created. By increasing the GST thickness from 360 nm to 540 nm, the peak absorption wavelength shifts approximately linearly from 9 μm to 13 μm, demonstrating the wavelength-selectivity of the device. In the simulation, a broad wavelength selectivity from 3 μm to 15 μm wavelength can be achieved (the permittivity of GST beyond 15 μm wavelength is unknown). Excellent agreement is obtained between the experimental data and the simulations. The slight difference can be attributed to light scattering loss induced by rough film surfaces.

To unveil the physics behind the sharp contrast between absorptivities of aGST-Au and cGST-Au samples, the specific absorptivity in each layer (GST layer and metal layer) is further investigated (Fig. 3a and b). Meanwhile, normalized electric field ($|E|$) and resistive loss ($Q$) distributions at resonant wavelengths for GST-Au samples are presented in Fig. 3c and Fig. 3d, respectively. It's obvious that the electric field intensity in the device decreases as light propagates into GST layers and it is quite weak at the GST-Au interface (Fig. 3c), demonstrating the π-phase shift owing to the reflection in metal and the $\lambda/4n$-thick GST layer. Absorption of light leads to heat generation, which can be characterized by resistive loss (also termed as "heat power volume density"). The resistive loss $Q$ is related to the imaginary part of material relative permittivity ($\varepsilon''$) and the intensity of electric field ($E$) by $Q = \pi c \varepsilon_0 \varepsilon''(\lambda)|E|^2/\lambda$, where $c$ is the light velocity in vacuum and $\varepsilon_0$ is the vacuum permittivity.[59] Owing to the transparent property of aGST, there is no resistive loss in the aGST layer even though its electric field is much stronger than that in the cGST layer. By comparison, large resistive losses occur in cGST layers for cGST-Au samples. The highly absorptive GST combined with resonant field enhancement contributes to nearly perfect absorption in cGST-Au samples. In both cases, there are resistive losses in the top surfaces of metal layers owing to its non-infinite imaginary part of the relative permittivity and residual electric field penetration into the metal.

The emitted spectra are measured by a Fourier transform infrared spectrometer (FTIR) equipped with a room-temperature doped triglycine sulfate (DTGS) detector. Thermal emission from the central areas (denoted by the red dotted circles in Fig. 4a) of the samples is sent into the FTIR for detection. Black soot is conventionally regarded as a perfect reference owing to its high wavelength-independent emissivity. Here the black soot reference is made by firing a rectangular stainless steel slice with a candle and its emissivity ($\varepsilon_{\text{ref}}(\lambda, T)$, where $\lambda$ is wavelength and $T$ is temperature) is assumed to be 0.97. With the measured emission spectra of the GST-Au samples ($S_{\text{sample}}(\lambda, T)$) and the black-soot reference ($S_{\text{ref}}(\lambda, T)$) (shown in Fig. S4a), the normalized emissivities

($\varepsilon_{sample}(\lambda, T)$) of the GST-Au samples can be derived from[12]:

$$\varepsilon_{sample}(\lambda, T) = \frac{S_{sample}(\lambda, T)}{S_{ref}(\lambda, T)} \cdot \varepsilon_{ref}(\lambda, T) \quad (1)$$

The derived emissivities of the GST-Au emitters with different GST thicknesses at 100 ℃ are presented in Fig. 4b. The cGST-Au emitter yields a notably high peak emissivity approaching the ideal blackbody maximum. The peak emissivities are 0.92, 0.95 and 0.97 for the GST thickness of 360 nm, 450 nm and 540 nm, respectively. The emissivities of aGST-Au emitters are low and the emission peaks are below 0.20. Therefore, this thermal emitter shows two distinct states: on-state with a high emissivity when GST is at the crystalline phase and off-state with a low emissivity when GST is at the amorphous phase. The extinction ratios and the differential emissivities between on and off states of the GST-Au emitters are presented in Fig. 4c and Fig. S4b, respectively. A maximum extinction ratio of 10.15 dB, 11.04 dB and 10.98 dB can be obtained for the GST thickness of 360 nm, 450 nm and 540 nm, respectively. The peak emission wavelength shifts longer with increasing cGST film thickness, showing promise as a wavelength-selective thermal emitter. Although the emissivity is measured directly while the absorptivity is characterized through transmission and reflection, excellent agreement between the experimental absorptivity (Fig. 2b) and emissivity (Fig. 4b) can be observed, verifying the Kirchoff's law of thermal radiation.

The visible and infrared photographs of the black-soot, the aGST-Au and cGST-Au emitters at 100 ℃ are presented in Fig. 5a. The temperature label of the infrared camera is based on the integration of received power from 8 μm to 14 μm. In the infrared photographs, it is obvious that the black soot shows strong emission while the aGST-Au emitter exhibits weak emission. In order to reveal angular dependence of emission, the oblique views of infrared photographs of the samples are presented in Fig. S5. The angular-dependent emitted power (from 8 μm to 14 μm) of the black soot and the cGST-Au emitter can be derived from the displayed temperature on the infrared camera. The normalized angular-dependent emitted power of the cGST-Au emitter with respect to the normally emitted power of the black soot is provided in Fig. 5b. As the angle increases, the thermal emission of both the black soot and the cGST-Au emitter decreases slowly, indicating that the thermal emission is robust with respect to the emission angle.

So far, we have mainly focused on switchable and wavelength-selective properties of the GST-Au emitters. Actually, the phase transition of GST between the amorphous phase and the crystalline phase is gradual rather than abrupt. During this phase transition, the GST film can be assumed to be at intermediate phases composed of different proportions of amorphous and crystalline molecules. By controlling different intermediate phases of GST, a tunable thermal emitter can be realized. This tunable property can be investigated by gradually increasing the sample temperature and simultaneously extracting the thermal emission spectrum. When the temperature is raised from 140 °C to 170 °C in 30 minutes with a step of 1 °C, the emissivity curves of the GST-Au emitter with a GST thickness of 450 nm are recorded and presented in Fig. 6a. The peak emissivity increases with temperature and rises sharply between 150 °C and 155 °C. The peak emissivity approaches the ideal blackbody maximum at 170 °C. The temperature-dependent emissivities at different wavelengths are explicitly presented in Fig. 6b, showing steep upslopes between 150 °C and 155 °C except for the 7 μm and 8 μm wavelengths. The temperature-dependent emissivity differences and extinction ratios in dB between the intermediate GST-Au samples and the aGST-Au sample are provided in Fig. S6. The relative refractive indices of GST during the intermediate phases are measured and presented in Fig. 6c. Both real and imaginary parts of the refractive index increase with temperature. The increase in the real part is primarily responsible for the resonance shift to longer wavelength while the increase in the imaginary part contributes to the increase in the emissivity. The intermediate phases of GST are also stable at room temperature and thereby no sustained external thermal consumption is needed to maintain a desired emissivity for the GST-Au emitter. The intermediate phase of GST is dominated by both annealing time and annealing temperature (Fig. S7). By controlling the annealing process, the crystallization level and the intermediate phase of GST can be correspondingly tailored, thus providing an avenue for tunable thermal emission.

In this paper, the crystallization process of GST is investigated by thermal annealing method while the reamorphization (phase transition from the crystalline phase back to the amorphous phase) is not applied owing to a low fusion point of the 120-nm-thick gold film. The thermal reamorphization of GST on a silicon wafer at a temperature above 640 ℃ is applied and the infrared properties are presented in Fig. S8, demonstrating the reamorphization ability of cGST. Since the Au film merely plays the role of a PEC, by replacing it with other metal films with higher fusion points (such as Ti, W and Mo), this emitter can also be reversely switchable from the on state back to the off state. The switching speed depends on the phase transition speed of GST. The phase transition process takes several seconds with thermal annealing method in our experiment. This timescale can be decreased by increasing annealing temperature (Fig. S7) or reducing the thermal substrate effect. Besides thermal annealing, the crystallization and reamorphization of GST can also be realized by proper laser pulses with high peak power[51, 60-63] or electrical stimulations.[47] The typical switching time for electrical stimulation is nanoseconds.[47] By stimulating the sample with laser pulses, the switching time can be nanoseconds[63] or even improved to femtoseconds.[61] So far, the emission property for the GST-Au emitter has been exclusively demonstrated in the 8-

14 μm atmospheric window. In fact, the cGST-Au samples have simultaneous switchable, tunable and wavelength-selective emission property in the 3-5 μm atmospheric window (shown in Fig. S9), demonstrating the capability of dual-band emission in both thermal atmospheric windows.

**Conclusions**

In summary, control over emissivity of zero-static-power mid-infrared thermal emitters consisting of a thermally stable GST layer and a metal layer is demonstrated. The GST film is free of sustained thermal consumption to preserve its phase. The metal film plays the role of a PEC, which not only halves the GST thickness, but also significantly enhances the resonant peak emissivity of the emitter. Broad wavelength-selectivity is realized simply by controlling the GST thickness. An emissivity approaching the ideal blackbody maximum is achieved when the GST is at the crystalline phase and a maximum extinction ratio of above 10 dB can be obtained between the two states (on state and off state) for the emitter. Besides, by controlling the intermediate phases composed of different proportions of amorphous and crystalline molecules of GST, the thermal emissivity can be continuously tunable. Moreover, since only layered films are involved in the fabrication, this thermal emitter presents the merits of large-area and lithography-free fabrication as well as design flexibility. Combining all these advantages, this switchable, tunable, wavelength-selective zero-static-power thermal emitter will pave the way towards the ultimate control of thermal emissivity in the field of fundamental science as well as for energy-harvesting and thermal control applications, including thermophotovoltaics, light sources, infrared imaging and radiative coolers.

**Acknowledgements**

This work is supported by the National Natural Science Foundation of China (Grant Nos. 61425023, 61575177, 61275030 and 61235007).

**Figure caption list**

FIG.1 (a) Atom distribution diagrams of the two phases (amorphous and crystalline) of GST. The red and blue dots denote the Ge/Sb atoms and Te atoms, respectively. (b) A 3D schematic of the switchable and tunable thermal emitter composed of a GST film on top of a gold film. (c) An SEM image of a cross-section of the fabricated thermal emitter.

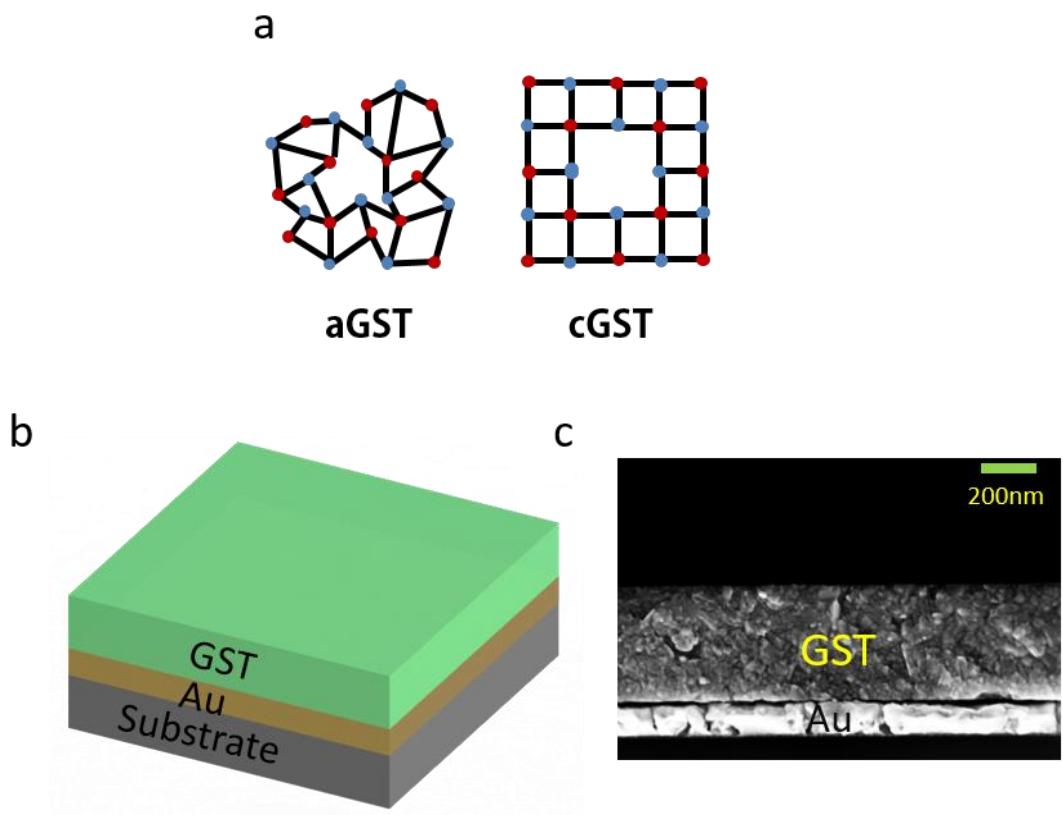

FIG.2 (a) and (b) are simulated and experimental normal-incident absorptivity of cGST-Au and aGST-Au samples in the mid-infrared with three different GST thicknesses (360 nm, 450 nm and 540 nm). The dashed and solid lines are for the aGST-Au and cGST-Au samples, respectively. (a) and (c) denote the aGST-Au and cGST-Au samples, respectively.

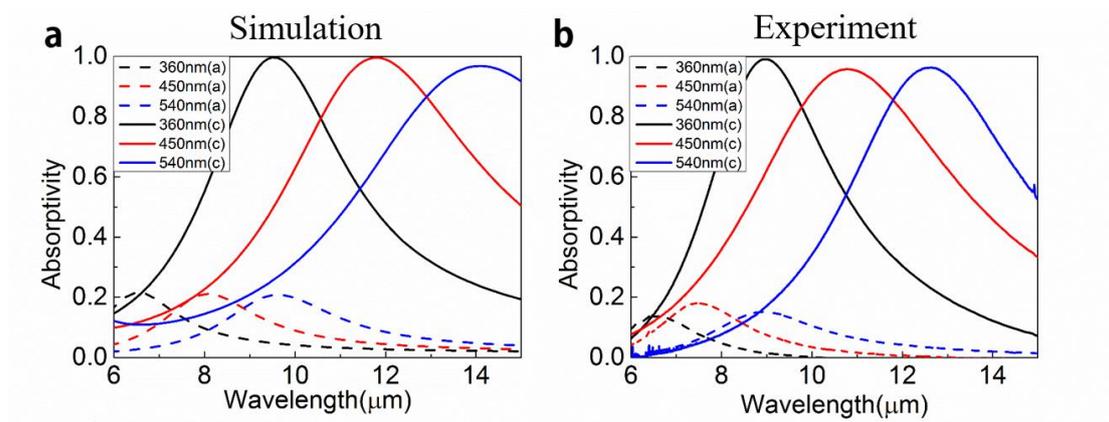

Fig.3 (a) and (b) Simulated absorptivity versus wavelength in specific layers (GST layer and metal layer) for the aGST-Au sample and the cGST-Au sample with different GST thicknesses (360 nm, 450 nm and 540 nm), respectively. m, a and c denote the metal layer, the aGST layer and the cGST layer, respectively. Simulated normalized resonant (c) electric field $|E|$ and (d) resistive loss $Q$ distribution in GST-Au samples with a 360-nm-thick GST film. The resonant wavelengths are 6.5 μm and 9.5 μm for aGST-Au and cGST-Au samples, respectively.

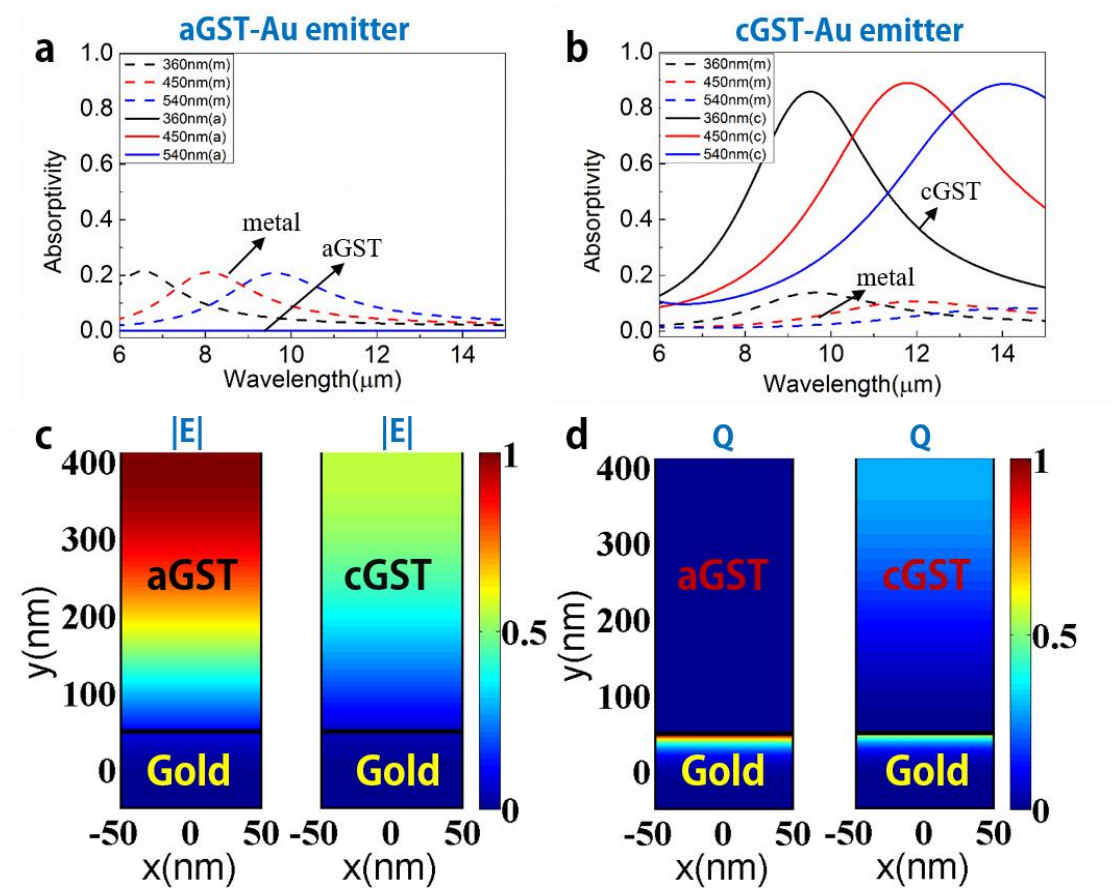

FIG.4 (a) Experimental photographs of the blacksoot (left) and the GST-Au emitter (right) on a temperature controller. (b) Thermal emissivity curves of GST-Au emitters with different GST thicknesses (360 nm, 450 nm and 540 nm) at 100 ℃. (c) Extinction ratio (in dB) between on and off states of the GST-Au emitter with different GST thicknesses at 100 ℃

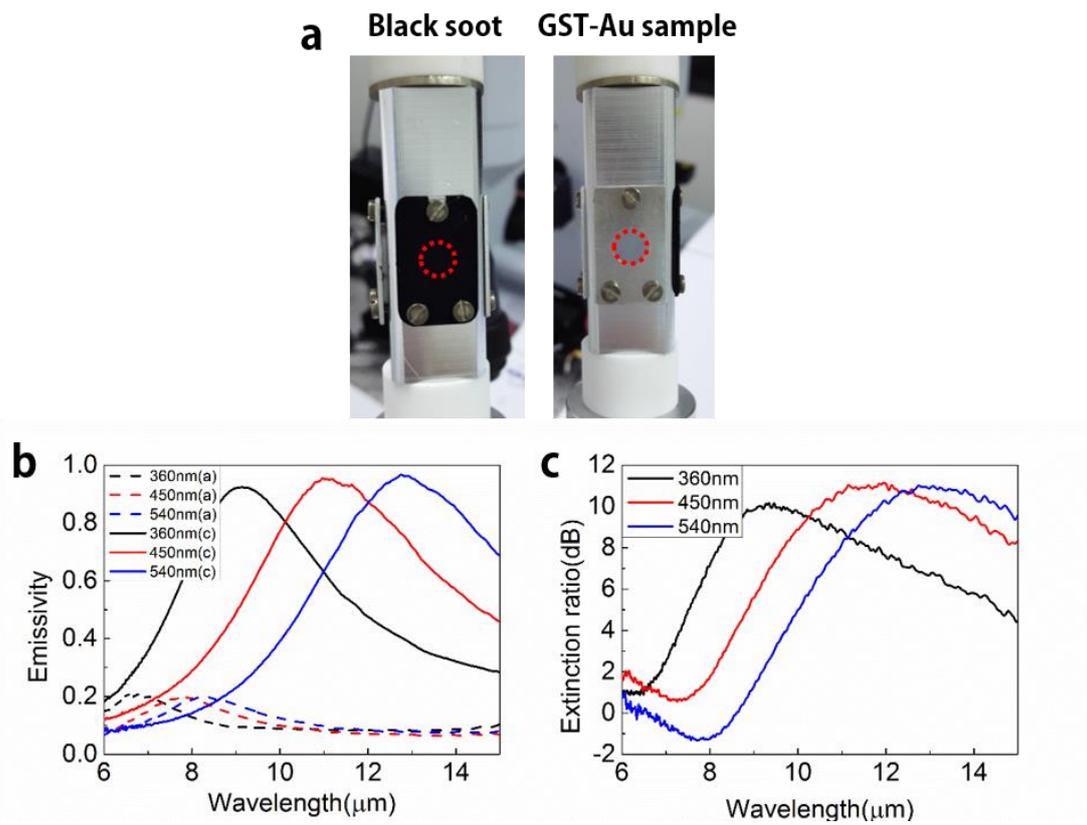

Fig.5 (a) Visible and infrared photographs of the black soot (left), the aGST-Au emitter (middle) and the cGST-Au emitter (right) at 100 ℃. (b) Measured emitted power from 0 ° to 60 ° for the black soot and the cGST-Au emitter. The GST thickness is 540 nm. The power is normalized to the emitted power of the black soot at 0 °.

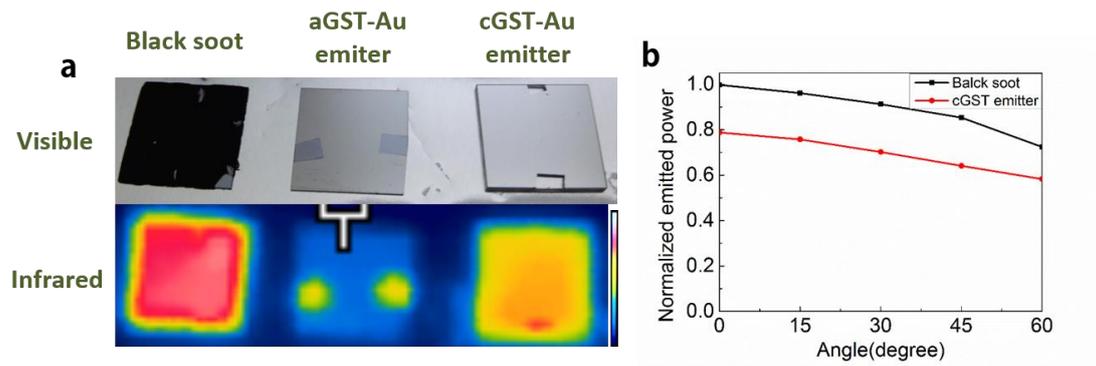

Fig. 6 (a) Tunable emissivity of the cGST-Au emitter in the mid-infrared with increasing temperature from 100 ℃ to 170 ℃. (b) Temperature response of the emissivity for the cGST-Au emitter at several typical wavelengths. (c) Measured refractive indices of GST at different temperature. n and k denotes real and imaginary parts of the refractive indices, respectively. The GST thickness is 450 nm.

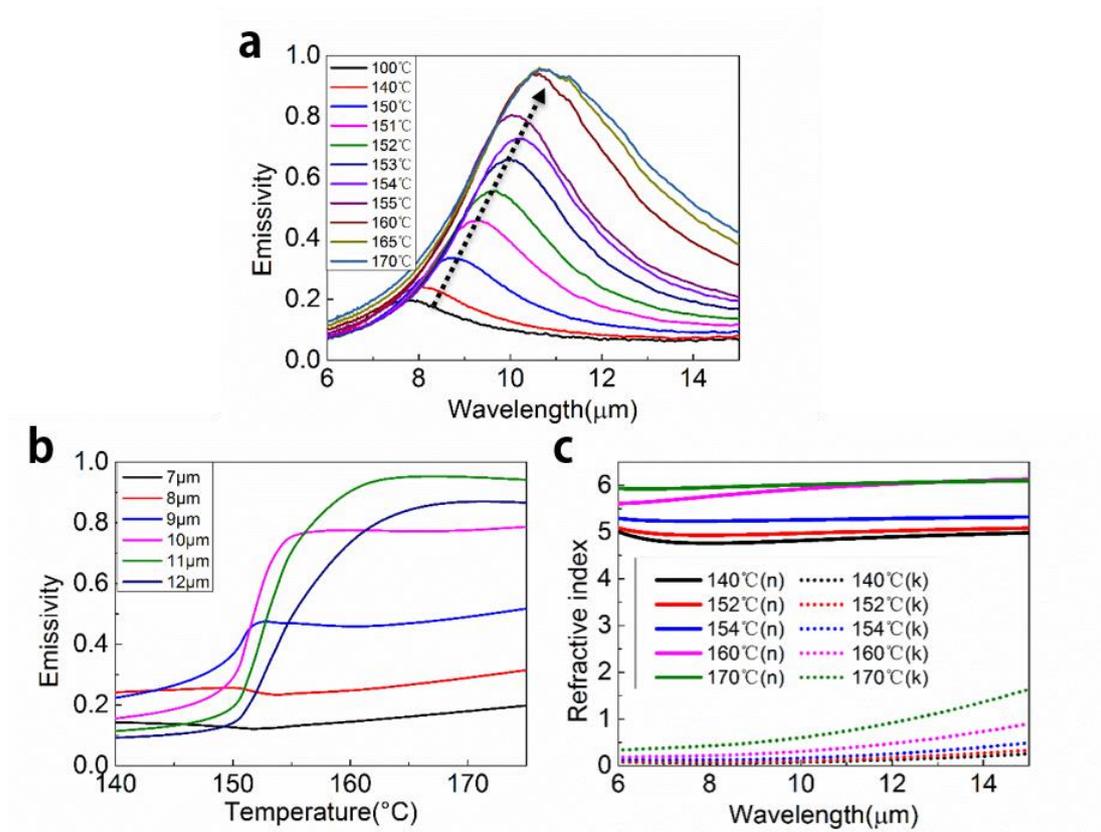